\documentclass{nature}
\usepackage{graphicx}
\usepackage[update,prepend]{epstopdf}
\usepackage{amsmath}
\usepackage{amssymb}
\usepackage{color}
\usepackage{ulem}
\usepackage{placeins}
\usepackage{multirow}
\usepackage[table]{xcolor}

\title{s-d coupling enhanced phonon anharmonicity in copper-based compounds}

\author{Kaike Yang$^{1,2}$, Huai Yang$^{1}$, Yujia Sun$^{1}$, Zhongming Wei$^{1}$, Jun Zhang$^{1}$, Jun-Wei Luo$^{1}$, Ping-Heng Tan$^{1}$, Shu-Shen Li$^{1}$, Su-Huai Wei$^{3}$ and Hui-Xiong Deng$^{1,*}$
\begin{affiliations}
\item State Key Laboratory of Superlattices and Microstructures, Institute of Semiconductors, Chinese Academy of Sciences, Beijing 100083, P. R. China 
 $\&$ Center of Materials Science and Optoelectronics Engineering, University of Chinese Academy of Sciences, Beijing 100049, P. R. China
\item Department of Physics, School of Physics and Electronics, Hunan Normal University, Changsha 410081, P. R. China
\item Beijing Computational Science Research Center, Beijing 100193, P. R. China \\
{\rm $^\star$Corresponding author, e-mail: hxdeng@semi.ac.cn (H.-X.D.)\\
\\
\bf{Materials with ultralow thermal conductivity are of great interest for efficient energy conversion and thermal barrier coating. Copper-based semiconductors such as copper chalcogenides and copper halides are known to possess extreme low thermal conductivity, whereas the fundamental origin of the low thermal conductivity observed in the copper-based materials remains elusive. Here, we reveal that {\it s-d} coupling induced giant phonon anharmonicity is the fundamental mechanism responsible for the ultralow thermal conductivity of copper compounds. The symmetry controlled strong coupling of high-lying occupied copper {\it 3d} orbital with the unoccupied {\it 4s} state under thermal vibration remarkably lowers the lattice potential barrier, which enhances anharmonic scattering between phonons. This understanding is confirmed by temperature-dependent Raman spectra measurements. Our study offers an insight at atomic level connecting electronic structures with phonon vibration modes, and thus sheds light on materials properties that rely on electron-phonon coupling, such as thermoelectricity and superconductivity.}
}
\end{affiliations}
}
\begin{document}
\maketitle
%\begin{abstract}
%\end{abstract}

%\section{Introduction.}
\begin{figure*}[t!]
\centering
\includegraphics[width=7cm]{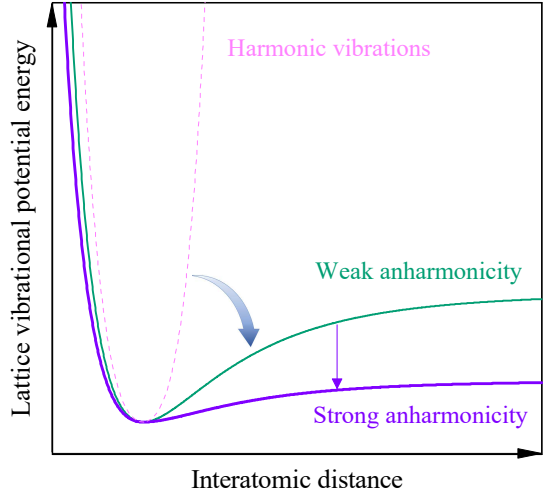}
\caption{Schematics of the anharmonicity in materials using a diatomic pair potential model, where the lattice vibrational potential energy as a function of interatomic distance is displayed, whereas dashed line indicates harmonic oscillation, cyan and violet curves are for anharmonic vibrations. The strength of the anharmonicity is quantified by the deviation of the potential energy from the harmonic one. In general, when atom is off equilibrium, the slower increase of the potential energy, the larger of the anharmonicity is in the system.}
\label{Model}
\end{figure*}

Copper-based materials often exhibit some unusual physical properties such as superconductivity\cite{Giustino2008,Bednorz1986}, superionicity\cite{Niedziela2019}, semimetallicity\cite{Tang2016}, antiferromagnetism\cite{Wadley2016}, transparent conductivity\cite{Nie2002}, and outstanding thermoelectricity\cite{Shi2012}, which are of particular importance in the fields of either condensed matter or energy engineering. For instance, thermoelectric devices that transform waste heat into useful electricity require building blocks with ultralow thermal conductivity ($\kappa$)\cite{Zhao2014,He2017,Mukhopadhyay2017} as well as high electrical conductivity. Liu $\it et~al.$ has reported experimentally that copper chalcogenides possess very low thermal conductivity with $\kappa<$1~$\rm W m^{-1} K^{-1}$ at room temperature\cite{Shi2012}. The lattice thermal conductivities of, e.g., $\rm Cu_2 S$\cite{He2014}, $\rm Cu_2 Se$\cite{Shi2012,Byeon2019}, CuCl\cite{Mukhopadhyay2018}, $\rm CuBiS_2$ and $\rm CuMX_2$ (M=Sb, Cr, and X=S, Se)\cite{Niedziela2019,Du2017,Zhang2016,Feng2017} are approximately two-orders of magnitude smaller than their adjacent compounds like GaP, GaAs, ZnS, ZnSe and NaCl in the periodic table\cite{Spitzer1970} and, interestingly, one order smaller than the leading thermoelectric materials such as PbTe\cite{Delaire2011}. The unusual intrinsic low thermal conductivity endows copper compounds as promising candidates for thermoelectric devices.

Traditionally, lower thermal conductivities are observed in materials with complex structures, heavy atoms or disordered arrangements in the unit cell\cite{Slack1973,Chiritescu2007,Snyder2008,Christensen2008,Poudel2008,Bera2010,Biswas2012,Voneshen2013,Katre2017,BLi2018,Morelli2008} in order to achieve small phonon group velocity and short mean free path. These traditional criteria clearly are unable to judge the thermal conductivity of copper-based compounds because they are all crystalline semiconductors with simple periodic structure and composed by relatively light chemical elements such as CuCl or $\rm Cu_2 S$. Liu $\it et~al.$ ascribed the low $\kappa$ of copper-based materials to the copper ion's liquid-like vibrations, which result in strong phonon scattering and some of the vibrational modes are completely suppressed\cite{Shi2012,He2014}. However, understanding the physical origin of the liquid-like vibrations remains ambiguous up to now, which is crucial for thermal management applications. Regarding the lattice dynamic theory\cite{Ziman1960,Mahan2000}, atomic vibrations in solids suffer from intrinsic scattering processes mainly by geometric boundary or scattering by other phonons. If the forces between atoms were purely harmonic, there would be no mechanism for collisions between different phonons. Therefore, the phonon scattering effects that are responsible for the low thermal conductivity here are mainly due to the large anharmonicity of the interatomic potential energy\cite{Slack1982,Mukhopadhyay2017,Shi2012,He2014}. In general, the anharmonicity is quantified by the deviation of the potential energy from the harmonic one, and the slower increase of the potential energy, the larger of the anharmonicity is in the system (see Fig.~1). Thereby, uncovering the nature of the potential energy in heated materials is the key. As for semiconductors, in addition to the ion-involved Coulomb interactions, the electronic coupling contributes significantly to the potential energy. For example, in PbTe, it has been demonstrated that the electronic configuration coined as resonant bonding is the fundamental mechanism of the low thermal conductivity\cite{Delaire2011,Lee2014,Li2015,Shportko2008}. However, for copper-based semiconductors, the correlation between electronic states and liquid-like copper sublattice vibrations (thus anharmonic phonon scattering) has not been established.

In this work, we reveal that the symmetry controlled strong {\it s-d} coupling is fundamentally responsible for the giant phonon anharmonicity in copper-based semiconductors. Cu has a highest occupied {\it 3d} orbital relative to other elements, yielding an extremely strong coupling between the high-lying copper occupied {\it 3d} state and the lowest conduction band state (an admixture of anion's {\it s} and copper's {\it 4s} orbitals), when atoms vibrate away from equilibrium. The remarkable red shift of phonon vibrational frequency with increasing temperature observed in the Raman spectra measurements experimentally confirms our theoretical expectation. This study reshapes the understanding of heat conduction in copper-based compounds and sheds light on the materials design that depends on engineering electron-phonon coupling.

%\section{Results.}
\begin{figure*}[t!]
\centering
\includegraphics[width=12cm]{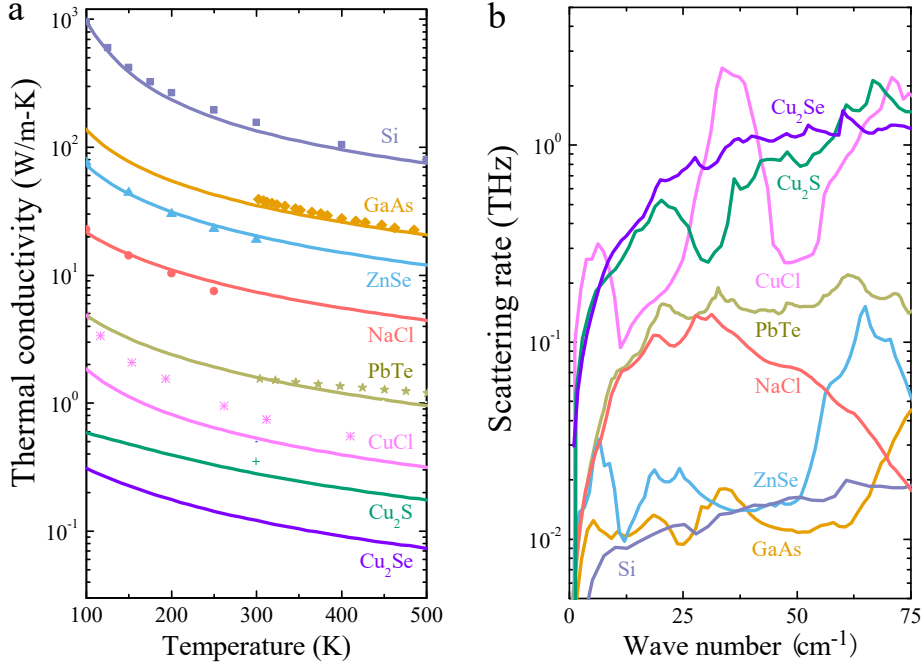}
\caption{Evaluation of lattice thermal conductivity and anharmonic phonon scattering. $\textbf{a}$, First-principles calculated thermal conductivity $\kappa$ (solid curves) as a function of temperature for $\rm Cu_2 Se$, $\rm Cu_2 S$, CuCl, PbTe, NaCl, ZnSe, GaAs and Si, while all experimental data (sparse dots) are taken from Refs.~[\citeonline{Slack1982,Steigmeier1966,Slack1972,Glassbrenner1964,Lu2018,Hakansson1986}], when available. For each material, we choose same color but different symbol to distinguish theoretical and experimental results. $\textbf{b}$, Phonon scattering rate $\gamma$ as a function of wave number for all investigated materials, where we have taken their averaged value of acoustic and optic branches as defined in Eq.~(\ref{meanrate}).}
\label{Conductivity}
\end{figure*}

Fig.~2a shows the comparison of the intrinsic lattice thermal conductivity of several prototypical crystalline semiconductors as a function of temperature. We calculate the temperature-dependent lattice thermal conductivity using the first-principles density functional theory (DFT) combined with the Boltzmann transport equation approach as implemented in VASP and Phono3py packages\cite{Dreizler1990,Kohn1965,Togo2015,Paulatto2015} (computational details are given in Methods and Supplementary Materials). We take $\rm Cu_2 S$, $\rm Cu_2 Se$ and CuCl as examples of copper-based compounds, and show their $\kappa(T)$ (solid line) in comparison with the results of Si, GaAs, ZnSe, NaCl and PbTe. We also show their corresponding experimental data (dots)\cite{Slack1982,Steigmeier1966,Slack1972,Glassbrenner1964,Lu2018,Hakansson1986,Spitzer1970}, when available. It exhibits that theoretical result is in good agreement with experimental data, which is particular true for groups IV, III-V and IIB-VI zinc-blende semiconductors. As temperature rises, the thermal conductivity decreases as expected due to enhanced phonon scattering effect. Moving from Si passing through GaAs, ZnSe, and NaCl, to PbTe, the thermal conductivity decreases monotonically at investigated temperature range. Interestingly, we find that the copper-based compounds exhibit very low thermal conductivity compared to traditional semiconductor materials, no matter how light their chemical component is. The lattice thermal conductivity of $\rm Cu_2 S$ and $\rm Cu_2 Se$ is nearly two orders magnitude smaller than that of adjacent GaAs and ZnSe in periodic table and one order smaller than that of PbTe.

In thermoelectric materials, Lee $\it et~al.$ ascribed the low $\kappa$ of PbTe to the large softening of transverse optic (TO) phonon modes which lead to strong scattering with acoustic phonons\cite{Delaire2011,Lee2014,Li2015,Shportko2008}. In Fig.~2b we show the phonon scattering rates as a function of wave number for all investigated materials. We find that copper-based semiconductors of $\rm Cu_2 S$, $\rm Cu_2 Se$ and CuCl exhibit the largest anharmonic scattering rates. Therefore, using phonon softening model, it is difficult to explain why CuCl has lower thermal conductivity (larger scattering) than PbTe, because the softening of TO modes in CuCl is weaker than that of PbTe (Fig.~S1). Although understanding the temperature dependence of thermal conductivity also requires taking into account changes both in the heat capacity of solids and in the propagation velocities of quasiparticles (Eq.~(\ref{LTC})), the phonon scattering effect dominates the behavior of $\kappa$ of copper compounds between 100 and 500~K compared to other materials (Fig.~S2).

%\section{s-d coupling mechanism.}
\begin{figure*}[t!]
\centering
\includegraphics[width=12cm]{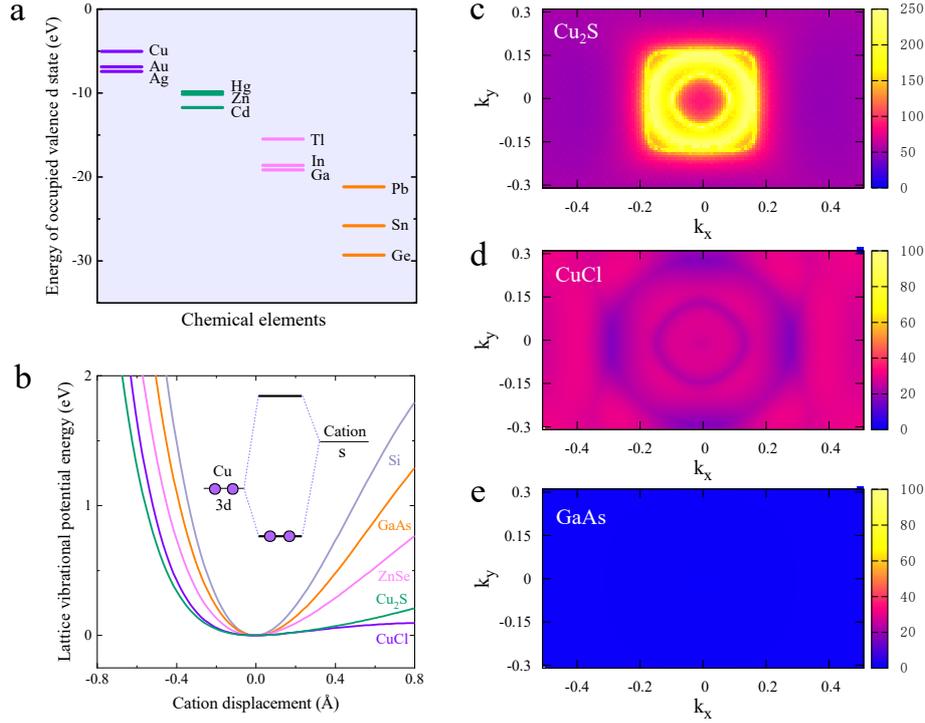}
\caption{Origin of the giant phonon anharmonicity in copper compounds. $\textbf{a}$, Chemical trend of the fully occupied valence {\it d} orbital energy level of elements in the Periodic Table, where all are referred to the same zero energy level at infinity. Obviously, Cu owns the highest valence {\it d} orbital energy level among elements with fully occupied valence {\it d} shell. $\textbf{b}$, First-principles calculated vibrational potential energy as a function of atomic displacement, where the potential energies at equilibrium are set as zero for comparison. We move cations along bond-stretching direction, i.e., Cu, Zn, Ga, or Si,in the copper compounds, ZnSe, GaAs, and Si, respectively (Fig.~S4). Clearly, copper-based compounds exhibit the lowest vibrational potential energy compared to traditional semiconductor materials at a given interatomic spacing. In the inset, we show schematically the band coupling between copper's {\it 3d} and {\it 4s} states. $\textbf{c}$, $\textbf{d}$ and $\textbf{e}$, The vibrational frequency shifts for $\rm Cu_2 S$, CuCl and GaAs, respectively, where $k_z=0$. We defined $\overline{\Delta\omega}=\frac{1}{N_{\nu}} \sum_{\nu}|\omega_{\mathbf{q}\nu}({\rm 300~K})-\omega_{\mathbf{q}\nu}({\rm 0})|$ to characterize the energy changes of phonon branches due to anharmonicity.}
\label{Anharmonicity}
\end{figure*}

To uncover the underlying mechanism why copper-based compounds have the low thermal conductivity, we analyzed the electronic structures and phonon vibrational properties. Compared to traditional semiconductors, the most prominent feature of copper compounds is the high-lying occupied {\it d} orbitals in energy (Fig.~S3). In zinc-blende structure with $T_d$ symmetry point group, the occupied {\it d} orbitals transform to irreducible representations of two-fold degenerate $E$ and three-fold degenerate $T_2$, respectively. Whereas, the high-lying unoccupied {\it s} orbital transforms as $A_1$. Therefore, there is no common symmetry between occupied {\it d} orbitals and unoccupied {\it s} orbital and their coupling is forbidden by symmetry. However, thermal vibrations reduce the crystal symmetry to $C_1$, in which the coupling between occupied {\it d} orbitals and unoccupied {\it s} orbital is allowed since {\it d} orbitals transform to $A$ and {\it s} to $A$. Such {\it s-d} coupling pushes the occupied {\it d} orbital down in energy and makes total energy smaller, which lowers the potential barrier of atomic vibrations. The strength of this allowed {\it s-d} coupling is proportional to the overlap of their wave functions but inversely proportional to the energy separation between occupied {\it d} orbitals and unoccupied {\it s} orbital. Thus, the anharmonicity is expected to enhance as the occupied {\it d} orbital energy increases.

Fig.~3a shows that copper owns the highest occupied {\it 3d} orbital in energy among all elements with fully occupied valence {\it d} states. The atomic energy of {\it 3d} electrons in copper is even higher than that of the outermost valence {\it p} states in chalcogen and halogen elements, but close to its own {\it 4s} state (Table~S1). In copper-based semiconductors such as copper halides, the copper's {\it 3d} orbital dominates the valence band maximum (VBM) state and is below the {\it s}-like conduction band minimum (CBM) by a band gap (about 3-4 eV) (Fig.~S3)\cite{Wei1993,Deng2016,Yang2019}. The small energy difference between copper's {\it 3d} and {\it 4s} states enhances their coupling which lowers the lattice vibrational potential energy. This is consistent with our first-principles calculations as shown in Fig.~3b, where we investigated the lattice vibrational potential energy of solids as a function of displacement of the atoms away from their equilibrium. For considered materials, we displace cations (i.e., Cu, Zn, Ga or Si) in the corresponding compounds along bond-stretching direction (see Fig.~S4). In the inset of Fig.~3b, we show how {\it s-d} coupling effect reduces the system potential energy and thus enhances the phonon anharmonicity. As moving from Cu to Zn and Ga, their occupied {\it 3d} orbital is getting deeper, i.e., away from the VBM (Table~S1), so that contributing less and less to the valence band edges and their {\it s-d} coupling strength is getting weak. It is found that GaAs and ZnSe have vibrational potential energy much higher than that of copper compounds such as $\rm Cu_2 S$ and CuCl, whereas Si exhibits the highest potential energy among them due to absence of the {\it s-d} coupling.

Although coupling between copper's unoccupied {\it 4s} orbital and anion occupied {\it p} state occurs in the copper-based compounds due to symmetry reduction, this effect is weak because (i) the unoccupied {\it 4s} state from cation and occupied {\it p} state from anion are separated in space, different from the {\it s-d} coupling in which both are from same copper ions; (ii) the energy difference between occupied {\it p} state of anion and unoccupied {\it s} state of copper is larger than that of copper's {\it 3d} and {\it 4s} orbitals, leading to the weak coupling strength of the former than the latter.

%\section{Analysis.}
\begin{figure*}[t!]
\centering
\includegraphics[width=12cm]{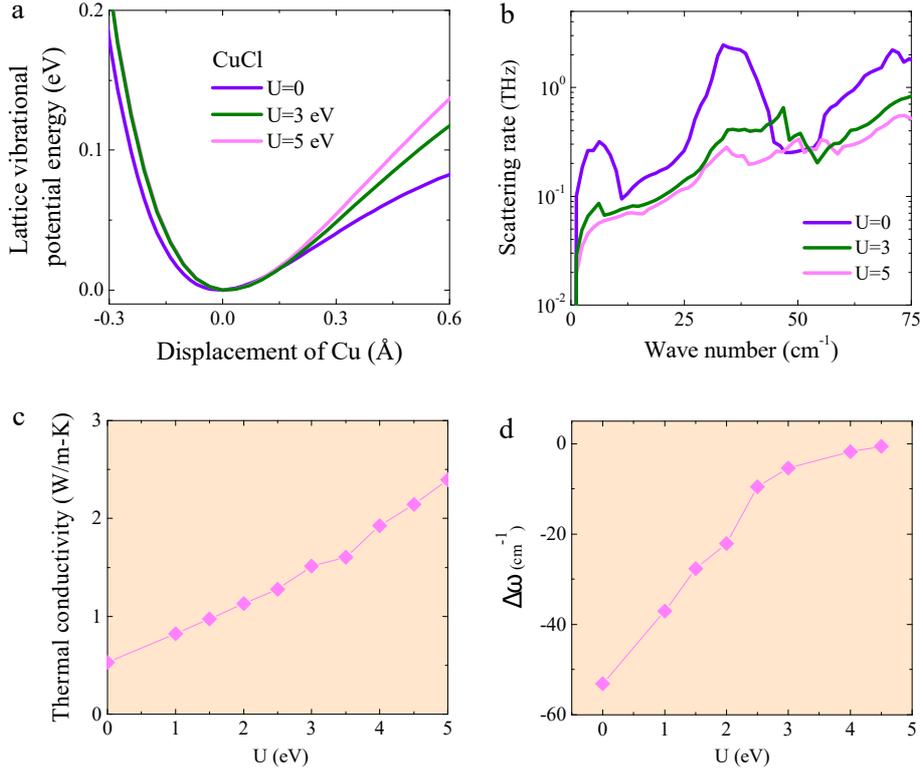}
\caption{Artificial modulation of s-d coupling effect in the thermal transport of copper-based semiconductors. $\textbf{a}$, Lattice vibrational potential energy of CuCl as a function of copper's displacement along the bond-stretching direction (Fig.~S4) under different Coulomb repulsive energy U, which is applied on Cu {\it 3d} orbitals. $\textbf{b}$ Scattering rate, and $\textbf{c}$ thermal conductivity of CuCl under different U. $\textbf{d}$, Anharmonicity caused frequency shift $\Delta\omega_{\mathbf{q}\nu}$ of optic phonon modes in CuCl between T=300 and 0 K at the center of the Brillouin zone.}
\label{DFTUcalculations}
\end{figure*}

In Figs.~3c and 3d, we investigate the lattice vibrational frequency shifts of $\rm Cu_2 S$ and CuCl due to anharmonic phonon interactions, respectively. To do so, we are employing the density functional perturbation theory as implemented in the Quantum-Espresso package to calculate the temperature dependent phonon frequency, $\Delta\omega_{\mathbf{q}\nu}(T)$, ($\Delta\omega_{\mathbf{q}\nu}(T)=\omega_{\mathbf{q}\nu}(T)-\omega_{\mathbf{q}\nu}(0)$). To reveal the rate of change of phonon energy in different materials, we defined $\overline{\Delta\omega}=\frac{1}{N_{\nu}} \sum_{\nu}|\omega_{\mathbf{q}\nu}(\rm 300 K)-\omega_{\mathbf{q}\nu}(0)|$ with $N_{\nu}$ being the number of phonon branches. As for $\rm Cu_2 S$, we find notable value of $\overline{\Delta\omega}$ near the center of Brillouin zone (Fig.~3c and Fig.~S5), whereas for CuCl, the value is reduced in magnitude (Fig.~3d and Fig.~S6). For comparison, in Fig.~3e, we have also presented $\overline{\Delta\omega}$ of GaAs. Interestingly, we find that the frequency shift of both $\rm Cu_2 S$ and CuCl is remarkable larger than that of GaAs, which has almost zero frequency changes until room temperature due to weak anharmonicity. The large temperature-dependent frequency shift of phonon modes confirms the strong anharmomic lattice dynamics in copper compounds.

Thus far, we have illustrated clearly the origin of the giant phonon anharmonicity in the copper-based semiconductor materials. To further verify our proposed model, in Fig.~4 we performed an artificial modulation by applying an advanced functional, i.e., DFT combined with the Hubbard U model\cite{Hubbard1963,Gao2018} calculations to adjust the {\it s-d} coupling strength to check the changes in the lattice thermal conductivity of copper compounds. According to the above discussion, it is expected that the lattice thermal conductivity will increase if we reduce the s-d coupling effect. Hence, we added a positive Coulomb repulsive energy, U, on Cu {\it 3d} orbital, which can lower the energy level of copper's occupied {\it 3d} orbitals. In this way, we can control the strength of {\it s-d} coupling in the copper-based compounds. As displayed in Fig.~4a, we find that the lattice vibrational potential energy of CuCl increases as U rises, due to the reduced {\it s-d} coupling strength resulting from the lowering of Cu {\it 3d} energy level and accompanied wave function localization\cite{Wei1993,Deng2016,Yang2019} (Fig.~S7). Accordingly, both the scattering rate shown in Fig.~4b and the thermal conductivity shown in Fig.~4c increases due to the reduced anharmonic phonon interactions. Moreover, in Fig.~4d, the calculated vibrational frequency shift of optic phonons at the center of Brillouin zone for CuCl at room temperature compared to low temperature decreases as the copper's {\it 3d} orbital moves away from the VBM. This result further supports that the {\it s-d} coupling is the root of the low thermal conductivity in copper-based materials.

%\section{Experiments.}
\begin{figure*}[t!]
\centering
\includegraphics[width=16cm]{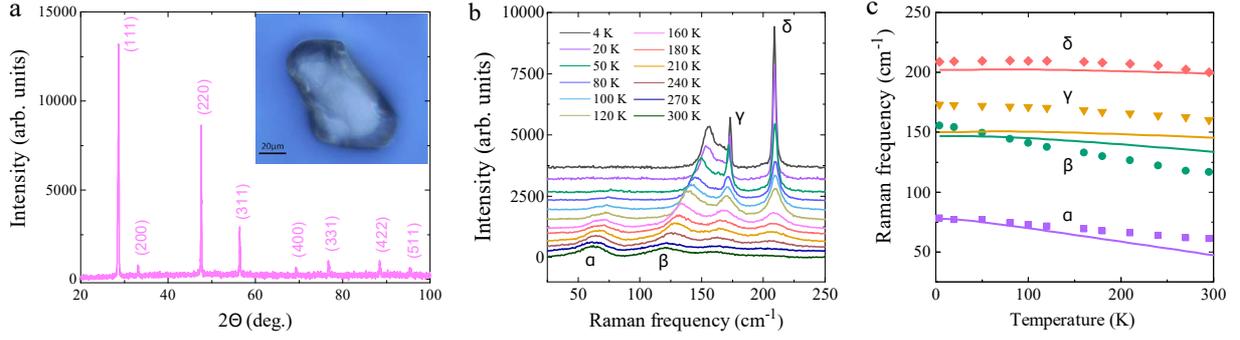}
\caption{Experimental evidence of phonon anharmonicity in copper-based materials. $\textbf{a}$, Wide angle X-ray diffraction patterns were obtained with an ultima IV diffractometer equipped with Cu $\rm K_\alpha$ radiation (40kV, 40mA) over the angle range 20 to 100$\rm ^o$. The XRD spectrum confirms that the CuCl crystal is single phase zinc-blende structure. The inset is the optical microscopy image of CuCl single crystal under a 100$\times$ objective lens. $\textbf{b}$, Raman spectra of CuCl single crystal in temperature range from 4~K to 300~K. From low to high frequency, as Potts $\it et~al.$ pointed out earlier\cite{Potts1974,Fukumoto1973}, $\alpha$ peak is contributed by the combination of longitudinal and transverse acoustic modes, $\beta$ and $\gamma$ bands are mainly from the transverse optic phonon modes, whereas $\delta$ band is due to the vibrations of longitudinal optic modes, respectively. $\textbf{c}$, Raman frequency shift of the four peaks labeled in ($\textbf{b}$). To compare with experimental results, we carried out first-principles calculations of the lattice vibrational frequency at finite temperature, which taking into account the anharmonic phonon interactions (solid lines).}
\label{Ramanmeasurements}
\end{figure*}

To verify the large anharmonicity of copper-based compounds, we also perform temperature-dependent Raman spectra measurements of copper chloride as an example. Fig.~5a shows the X-ray diffraction (XRD) pattern of single crystal CuCl, where sample synthesis and characterization are illustrated in detail in the Methods. We find that XRD shows clearly all the characteristic peaks of zinc-blende structure with space group $\rm F\bar{4}3m$, which confirms the sample quality of CuCl single crystal. In Fig.~5b we present Raman measured vibrational spectra in a wide temperature range from 4~K to 300~K. Interestingly, four pronounced lattice vibrational peaks labeled by $\alpha$, $\beta$, $\gamma$ and $\delta$ corresponding to the phonon energy of $\omega\approx$ ${\rm 65~cm^{-1}}$, ${\rm 155~cm^{-1}}$, ${\rm 166~cm^{-1}}$ and ${\rm 209~cm^{-1}}$ are identified, respectively. Earlier, Potts $\it et~al.$ pointed out that the symmetric $\delta$ peak is from the zone-center vibrations of longitudinal optic modes, whereas $\alpha$ band is a combination of the longitudinal and transverse acoustic modes\cite{Potts1974,Fukumoto1973}; as to the $\beta$ and $\gamma$ bands, it is found that they are dominated by the transverse optic phonon modes vibrations, respectively.

In Fig.~5c we show the Raman frequency shift of the four vibrational peaks indicated in 4b. As temperature rises, all peaks shift to the lower frequency regime, and a remarkable energy reduction for $\alpha$ and $\beta$ bands occurs. This suggests the significant softening of the phonon modes and confirms the giant phonon anharmonicity in copper-based compounds. Our first-principles simulations (solid lines) are consistent with the experimental observations.

%\section{Discussion.}
Besides the binary copper compounds, recently, it was found that few ternary copper-based materials also exhibit very low thermal conductivity (Table~S2), such as $\rm CuBiS_2$\cite{Feng2017}, $\rm CuSbX_2$\cite{Du2017,Zhang2016} and $\rm CuCrX_2$ (X=S or Se)\cite{Niedziela2019,Bhattacharya2013,Xia2020}. Particularly, Niedziela $\it et~al.$ demonstrated that $\rm CuCrSe_2$ has liquid-like thermal diffusive behavior at a relative high temperature, causing a large lattice anharmonicity due to copper ions dominated phonon modes breaking down in the low-energy regime\cite{Niedziela2019}. Applying above theory, it is easy to understand this behavior, which is caused mainly by the strong band coupling between copper ion's unoccupied {\it 4s} and occupied {\it 3d} states. This study indicates the prominent role of copper ions in the heat conduction in this class of materials.

%\section{Conclusions.}
Using combined first-principles calculations and experimental measurements, we investigated the fundamental origin of the ultralow thermal conductivity observed in the copper-based semiconductor materials. We revealed that the symmetry controlled coupling between copper's occupied {\it 3d} and unoccupied {\it 4s} orbitals lowers the lattice vibrational potential and promotes interactions of the collective oscillations. Therefore, the anharmonic scattering between phonons is enhanced which suppresses the crystalline thermal conductivity in copper-based compounds. This study, connecting the electronic structures and phonon vibrational modes, offers a new insight to understand the behavior of thermal transport in semiconductor materials.

%% Put the bibliography here.
%\begin{thebibliography}
%\end{thebibliography}
\paragraph{References}
\bibliography{Anharmonicity}

\begin{Methods}
\paragraph{First-principles determination of lattice thermal conductivity and scattering rate.}
After fully relaxation of the atomic positions in a solid, we employed the first-principles density functional theory calculations as implemented in the VASP package combined with the Boltzmann transport equation approach implemented in the Phono3Py program to investigate the lattice vibrational properties of a material\cite{Dreizler1990,Kohn1965}, whereas computational parameter settings are illustrated in Supplementary Materials. Therefore, the phonon-dominated lattice thermal conductivity in semiconductors within relaxation time approximation is expressed as
\begin{equation}
\kappa=\frac{1}{N_{\mathbf{q}}\Omega} \sum_{\mathbf{q}\nu} \mathbf{v}_{\mathbf{q}\nu}\otimes\mathbf{v}_{\mathbf{q}\nu} c_{\mathbf{q}\nu} \tau_{\mathbf{q}\nu}.
\label{LTC}
\end{equation}
Here, $N_{\mathbf{q}}$ is the number of sampling $\mathbf{q}$ points in the Brillouin zone, $\Omega$ the volume of investigated system, $\mathbf{v}_{\mathbf{q}\nu}$ the group velocity of phonon with wave vector $\mathbf{q}$ and branch $\nu$, and $c_{\mathbf{q}\nu}$ the heat capacity. The mode-dependent phonon scattering rate is written in the form of\cite{Togo2015}
\begin{equation}
\begin{split}
&\gamma_{\mathbf{q}\nu}(\omega)=\frac{1}{2\tau_{\mathbf{q}\nu}(\omega)}= \\
&\frac{36\pi}{\hbar^2}\sum_{\mathbf{q}'\nu'}\sum_{\mathbf{q}''\nu''}|V^{(3)}_{\mathbf{q}\nu,\mathbf{q}'\nu',\mathbf{q}''\nu''}|^2 \delta(\mathbf{q}+\mathbf{q}'+\mathbf{q}''-\mathbf{G}) \times \\
&\{(n_{\mathbf{q}'\nu'}+n_{\mathbf{q}''\nu''}+1) \delta(\omega-\omega_{\mathbf{q}'\nu'}-\omega_{\mathbf{q}''\nu''})+ \\
&~~(n_{\mathbf{q}'\nu'}-n_{\mathbf{q}''\nu''}) [\delta(\omega+\omega_{\mathbf{q}'\nu'}-\omega_{\mathbf{q}''\nu''})- \\ &~~~\delta(\omega-\omega_{\mathbf{q}'\nu'}+\omega_{\mathbf{q}''\nu''})]\},
\end{split}
\label{Linewidth}
\end{equation}
where $\tau_{\mathbf{q}\nu}$ is the phonon lifetime, $\hbar$ the reduced Planck constant, $V^{(3)}_{\mathbf{q}\nu,\mathbf{q}'\nu',\mathbf{q}''\nu''}$ the third-order anharmonic matrix element, summation taking over all scattering processes in which a phonon with wave vector $\mathbf{q}$ decays into two phonons ($-\mathbf{q}'\nu')$ and ($-\mathbf{q}''\nu''$), or in which a phonon with wave vector $\mathbf{q}$ initially coalesces with ($\pm\mathbf{q}'\nu'$) and then emits ($\mp\mathbf{q}''\nu''$), while energy and momentum conservation of the process is guaranteed by the Dirac functions. One can distinguish between normal and umklapp scattering processes by choosing $\mathbf{q}'$ and $\mathbf{q}''$ when $\mathbf{q}$+$\mathbf{q}'$+$\mathbf{q}''$=0 or $\mathbf{G}$ with $\mathbf{G}$ being the reciprocal lattice vector, respectively. To qualify the distribution of phonon modes, it is convenient to define an averaged scattering rate as
\begin{equation}
\gamma(\omega)=\frac{1}{N_{\nu} N_{\mathbf{q}}} \sum_{\mathbf{q}\nu}\gamma_{\mathbf{q}\nu}(\omega),
\label{meanrate}
\end{equation}
where $N_{\nu}$ is the total number of acoustic and optic phonon branches.

\paragraph{Vibrational frequency shift due to phonon anharmonicity.}
To calculate the frequency shift due to anharmonic phonon interactions, we employed first-principles density functional perturbation theory as implemented in the Quantum-Espresso package\cite{Giannozzi2009}. Therefore, the temperature-dependent lattice vibrational frequency is given by\cite{Paulatto2015} $\omega_{\mathbf{q}\nu}(T)=\omega_{\mathbf{q}\nu}(0)+\Delta\omega_{\mathbf{q}\nu}$, with
\begin{equation}
\begin{split}
&\Delta\omega_{\mathbf{q}\nu}= \\
&Re\bigg\{\frac{1}{\hbar}\sum_{\mathbf{q}'\nu'}\sum_{\mathbf{q}''\nu''}|V^{(3)}_{\mathbf{q}\nu,\mathbf{q}'\nu',\mathbf{q}''\nu''}|^2 \delta(\mathbf{q}+\mathbf{q}'+\mathbf{q}''-\mathbf{G}) \times \\
&~~~~~~\bigg[\frac{(\omega_{\mathbf{q}'\nu'}+\omega_{\mathbf{q}''\nu''})(n_{\mathbf{q}'\nu'}+n_{\mathbf{q}''\nu''}+1)}
{{(\omega+i\eta)^2-(\omega_{\mathbf{q}'\nu'}+\omega_{\mathbf{q}''\nu''})^2}}+ \\ &~~~~~~\frac{(\omega_{\mathbf{q}'\nu'}-\omega_{\mathbf{q}''\nu''})(n_{\mathbf{q}''\nu''}-n_{\mathbf{q}'\nu'})}
{{(\omega+i\eta)^2-(\omega_{\mathbf{q}'\nu'}-\omega_{\mathbf{q}''\nu''})^2}}\bigg]\bigg\},
\end{split}
\label{Frequencyshift}
\end{equation}
where $\omega_{\mathbf{q}\nu}(0)$ is the frequency at harmonic oscillations. $\eta$ is a small positive quantity added to avoid divergence in practical computations and $n_{\mathbf{q}\nu}$ the Bose-Einstein distribution function.

\paragraph{Sample synthesis and characterization.}
High-quality single crystal of CuCl with zinc-blende lattice was synthesized via chemical vapor transport technique technique. 1.9g CuCl powder (Alfa, 99.999\%) and 0.1g KCl powder (Alfa, 99.999\%) were placed in an ampoule. The KCl powder was used to lower the temperature below the phase transition point (407$\rm ^o C$) and to avoid forming undesired wurtzite CuCl crystals\cite{Kaifu1967}. The ampoule was evacuated to the lowest pressure attained in our lab ($10^5$ Torr) and sealed immediately. The synthesis was conducted in a horizontal two-zone tube furnace system. The growth temperature gradient was set to 350/450$\rm ^o C$ under heating 10 hours. Samples were left at the temperature difference for 10 days. The system was finally cooled down to room temperature at a 10$\rm ^o C$/h rate. The CuCl crystals picked up are transparent colorless with several millimeters long mainly concentrated in the bottle of the quartz tube. X-ray diffraction shows clearly all characteristic peaks of the single crystal CuCl with space group $\rm F\bar{4}3m$, which confirms the sample quality.

\paragraph{Temperature-dependent Raman measurements.}
The substrate we used was a p-type (100) Si wafer with 300 nm thick $\rm SiO_2$ layer. Raman measurements were undertaken in backscattering geometry with a Jobin-Yvon HR800 system equipped with a liquid-nitrogen-cooled charge-coupled detector. The Raman spectra were collected with a 50$\times$ long-working-distance objective lens (NA=0.5) and a 1800 lines/mm grating at low temperature measurements. The excitation wavelength for zinc-blende cubic CuCl is 442 nm from a conventional He-Cd laser and the typical laser power is about 0.8 mW to avoid laser heating on sample.
\end{Methods}
\\
%% Here is the endmatter stuff: Supplementary Info, etc.
\begin{Acknowledgements}
This work was supported by the National Natural Science Foundation of China (Grants No. 61922077, No. 11804333, No. 11704114, No. 11874347, No. 61121491, No. 61427901, No. 11634003 and No. U1930402), the National Key Research and Development Program of China (No. 2018YFB2200100), Beijing Science and Technology Committee (No. Z181100005118003). H.-X.D. was also supported by the Youth Innovation Promotion Association of Chinese Academy of Sciences (Grant No. 2017154).
\end{Acknowledgements}
\\
%% We put all additional figures at last.
%% If there are any tables, put here.
%% THE END
\end{document}